\documentclass[aps,amsfonts,amsmath,amssymb,bm
,twocolumn
]{revtex4}

\usepackage{graphicx,color}

\begin{document}

\title{An explanation for the glass-like anomaly in the \\ low-temperature
specific heat of incommensurate phases}

\author{A. Cano}
\email{andres.cano@uam.es}
\author{A. P. Levanyuk}
\email{levanyuk@uam.es}
\affiliation{Departamento de F\'\i sica de la Materia Condensada C-III,\\
Universidad Aut\'onoma de Madrid, E-28049 Madrid, Spain}
\date{\today}

\begin{abstract}
An explanation for the glass-like anomaly observed in the low-temperature specific heat of incommensurate phases is proposed. The key point of this explanation is the proper account for the phason damping when computing the thermodynamic magnitudes. The low-temperature specific heat of the incommensurate phases is discussed within three possible scenarios for the phason dynamics: no phason gap, static phason gap and a phason gap of dynamical origin. Existing NMR and inelastic scattering data indicate that these scenarios are possible in biphenyl, blue bronze $\rm K_{0.30}MoO_3$ and BCPS respectively. Estimates of the corresponding low-temperature specific heat are in reasonable agreement with the experiments.
\end{abstract}

\maketitle

Deviations from the Debye law observed in the low-temperature specific heat of incommensurate (IC) phases have attracted some attention \cite{Etrillard96,Etrillard97,Dahlhauser86,Odin01}. It is observed, in particular, a quasi-linear-in-$T$ contribution to this specific heat ($C$) at the lowest temperatures ($\lesssim 1\, \rm K$) and, in some compounds \cite{Etrillard96,Dahlhauser86,Odin01}, a maximum in $C(T)/T^3$ at temperatures $\sim 10\,\rm K$. It is expected that these peculiarities are connected to the low-energy excitations specific to IC phases, i.e. the phasons. Theoretical discussions try to relate these phasons and the mentioned maximum in $C(T)/T^3$ \cite{Etrillard96,Odin01}. As a result, this maximum is usually ascribed to existence of a gap in the corresponding phason spectrum. The origin of the quasi-linear-in-$T$ contribution, in contrast, has not been elucidated. Nevertheless, due to its similitude to that observed in glasses, this contribution inspire some speculations \cite{Odin01}. 

In this paper we show that these peculiarities in the low-temperature specific heat of IC phases have a natural explanation. This explanation is based on a specific feature of phasons that is overlooked in the above cited works: the phason damping. In fact it follows from general considerations that, even in perfect crystals, phasons will be overdamped for small enough wavevectors \cite{Golovko81}. The phason damping in perfect crystals has a strongly dependence on temperature and vanishes for $T=0$ \cite{Levanyuk97a}. Such a temperature-dependent damping, however, is not consistent with the NMR \cite{Liu85,Souza91} and the inelastic neutron scattering data \cite{Launois89,Ollivier98,Ollivier03} reported for the IC phases in which we are interested: the damping inferred from these experiments is twelve orders of magnitude stronger than the calculated for perfect crystals (see below). The phason damping in real crystals may then be due to defects. To the best of our knowledge, there is no theoretical calculations of such a defect-induced phason damping. Anyway, to our purposes, this damping can be estimated from the above mentioned experiments.

It is worth mentioning that, in some sense, phasons in IC phases are analogous to vortons in superconducting vortex lattices. They both represent acoustical vibrations of the corresponding IC structure that, unlike to the ordinary acustical vibrations, are overdamped for small enough wavevectors. In fact, it has already been found that vortons yield a linear-in-$T$ contribution to the low-temperature specific heat of the corresponding vortex lattice \cite{Bulaevskii93,Iengo98}. The key point of the theory we present below coincides with that of the mentioned works on vortex lattices: the proper account for the specific thermodynamic features of the damped oscillators (phasons in our case). Mention also that these specific features have been found relevant even for ``ordinary'' crystals: optical vibrations in real crystals remain damped due to defects down to zero temperature, so they may give an important contribution to the corresponding low-temperature specific heat \cite{Cano/cond-mat/0404063}. In this paper we shall concentrate on IC phases mainly because i) the smallness of the phason frequencies increases the importance of the phason damping and, consequently, of the corresponding contribution to the specific heat and ii) the existence of experimental data characterizing the phason dynamics makes it possible to estimate in order of magnitude.

Let us first consider, for the sake of illustration, the simple case in which the phason spectrum shows an \emph{static phason gap} (due to ``phase pinning,'' see e.g. Ref. \cite{Blinc85}). If the corresponding wavevector-dispersion is negligible, the phason branch becomes analogous to an optical one within the Einstein approximation. The corresponding contribution to the specific heat can then be computed from the specific heat of an harmonic oscillator $C_\text{osc}$. But to describe correctly the low-temperature regime in the IC phases, it is crucial to take into account the phason damping. This can be done by using the formalism of Refs. \cite{Bulaevskii93,Weiss}. As a result one obtains
\begin{align}
C_\text{osc} = k_B
\left[\sum_{i=1}^3 
\Big({\lambda_i\over \nu}\Big)^2
\psi'\Big({\lambda_i\over \nu}\Big)
-\Big({\omega_D \over \nu}\Big)^2
\psi'\Big({\omega_D\over \nu}\Big)-1\right],
\label{specific_oscill}\end{align}
where $\nu=2\pi k_BT/\hbar$, $\psi'(x)=d^2[\ln \Gamma(x)]/dx^2$ is the trigamma function, and the $\lambda$'s are the roots of the equation
\begin{align}
\lambda^3 + \omega_D\lambda^2 +(\omega_0^2+\gamma \omega_D)\lambda + \omega_D\omega_0^2 =0,
\end{align}
with $\omega_0$ the natural frequency of the oscillator and $\gamma $ the damping coefficient, i.e. the viscosity coefficient divided by the mass of the oscillator. Here $\omega_D^{-1}$ is a memory time associated with a Drude regularization of this viscosity damping (see, e.g., Ref. \cite{Weiss} and the Appendix). In further considerations the limit $\omega_D\to \infty$ is taken. In this limit, the regularized viscosity damping reduces to the ``ordinary'' one. 

\begin{figure}
\includegraphics[width=.4\textwidth,clip]{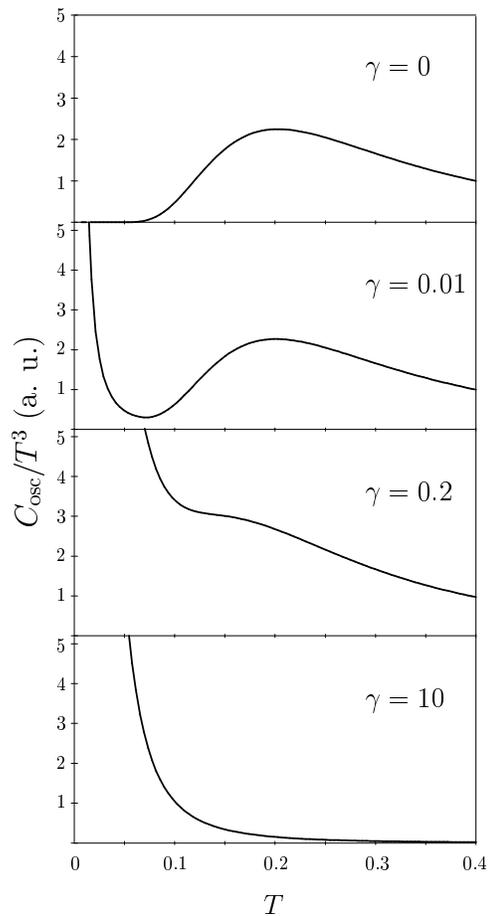}
\caption{\label{fig:1}Specific heat Eq. \eqref{specific_oscill} divided by $T^3$ of an harmonic oscillator for different dampings. Here units have been chosen such that $k_B=\hbar =1$, the natural frequency of the oscillator as $\omega_0=1$ and the Drude frequency (see, e.g., Ref. \cite{Weiss} and the Appendix below) $\omega_D=10^{4}$. A different scale in $C_\text{osc}/T^3$ is used in the plot for $\gamma =10$.}
\end{figure}

The dependence on temperature of the specific heat Eq. \eqref{specific_oscill} of a damped harmonic oscillator can be seen in Fig. \ref{fig:1} for different values of the damping coefficient $\gamma$. In the undamped case ($\gamma=0$), it is well known that this specific heat is exponentially small at low enough temperatures ($k_BT\ll\hbar \omega_0$). For any finite damping, however, the low-temperature asymptotic for the specific heat Eq. \eqref{specific_oscill} is linear-in-$T$:
\begin{align}C_\text{osc} \underset{T\to 0}{
\approx} {\pi k_B\over 3}{\gamma\over \omega_0}{k_B T\over \hbar \omega_0}.
\label{specif_one_osc}\end{align}
In overdamped case ($\gamma \gg \omega_0$) this expression is valid for $k_B T\ll \hbar \omega_0^2/\gamma$ while in underdamped one ($\gamma \ll \omega_0$), it does for $k_B T \ll \hbar \omega_0$. 

It is worth noticing the qualitative similitude, already at this level of  consideration, between this specific heat and that reported in IC phases \cite{Etrillard96,Etrillard97,Dahlhauser86,Odin01}. In particular in what concerns to the linear-in-$T$ dependence at the lowest temperatures, which is connected to the damping as we have seen. In the following we shall make more concrete considerations.

Let us then consider the Landau thermodynamic potential (see e.g. Ref. \cite{Ollivier98,Levanyuk97a})
\begin{align}
\phi = \phi_0+{a\over 2} (\eta_1^2\negthinspace+\eta_2^2)+{b\over 4} (\eta_1^2\negthinspace+\eta_2^2)^2\negthickspace
+{c\over 2}[(\nabla \eta_1)^2\negthinspace+(\nabla \eta_2)^2],
\label{landau}\end{align}
where $\eta_1$ and $\eta_2$ are the real and imaginary part of the complex order parameter $\eta =(\eta_1,\eta_2)$. In the IC phase ($a<0$), the equilibrium values can be taken such that $\eta_1^\text{(eq)}=(-a/b)^{1/2}$ and $\eta_2^\text{(eq)}=0$. Within the \emph{scheme of no phason gap}, which seems to be valid for biphenyl (see below), for small deviations of the order-parameter components from their corresponding equilibrium values: $\eta_i=\eta_i^\text{(eq)}+\eta_i'$, one has the following equations of motion (see e.g. Ref. \cite{Levanyuk97a}): 
\begin{subequations}\label{eq_mot_eta}\begin{align}
m \ddot \eta_1'+ \gamma_\eta  \dot \eta_1 -2a \eta_1'-c\nabla^2\eta_1'& = 0,
\label{eq_mot_eta_1}
\\
m \ddot \eta_2'+ \gamma_\eta  \dot \eta_2 -c\nabla^2\eta_2'&=0.
\label{eq_mot_eta_2}
\end{align}\end{subequations}
Hence $\eta_1$ is associated with longitudinal (amplitude) fluctuations while $\eta_2$ does with transverse (phase) ones. 

In Eq. \eqref{eq_mot_eta_2} it is implicit that, as stated above, for small enough wave-vectors phasons are overdamped due to the viscosity term $\gamma_\eta \dot \eta_2$. At low enough temperatures the phason contribution to the specific heat will be linear-in-$T$. From Eq. \eqref{specif_one_osc}, this contribution can be estimated as 
\begin{align}C \approx {k_m^3\over 6\pi}{k_B^2 T\over \hbar (v^2k_m^2/\gamma)}\label{low_T_specificheat_relax}\end{align}
for $k_B T/\hbar \ll \min (vk_m,v^2k_m^2/\gamma)$; where $v^2=c/m$, $\gamma=\gamma_\eta/m$ and $k_m$ represents the radius of the Brillouin zone. (Here it has been assumed that the damping is the same for whole phason branch.) This linear-in-$T$ contribution to the low-temperature specific heat of IC phases due to phasons will prevails over the Debye one (due to acoustic phonons) at temperatures $T \lesssim [\gamma /(vk_m)]\theta$, where $\theta= (\hbar/k_B)(V^3/v)^{1/2}k_m$ with $V$ the velocity of sound.

Let us mention that, in accordance with Eq. \eqref{eq_mot_eta_1}, amplitude fluctuations of the order parameter (amplitudons) are also damped. In consequence, they also give a linear-in-$T$ contribution to the specific heat at the lowest temperatures. However, this contribution is always smaller than phason one because, for a given wavevector, the amplitudon frequency is always greater than the phason one, the corresponding damping coefficient being similar in both cases [see Eqs. \eqref{eq_mot_eta} and \eqref{specif_one_osc}]. Indeed the amplitudon contribution to the low-temperature specific heat of the IC phases of interest is not essential to describe the corresponding experimental data, so we shall omit this contribution if further considerations. 

Let us now consider the \emph{scheme of a dynamic phason gap} \cite{Ollivier98}. In some compounds such as BCPS the coefficients of the equations of motion for the order-parameter may have some frequency dispersion. A relatively simple dispersion is described in the phenomenological model developed in Ref. \cite{Shapiro72} to reproduce the central peak in the high-temperature phase of strontium titanate. Such a dispersion can be understood as a result of the linear coupling between the order-parameter and a relaxational variable $\xi=(\xi_1,\xi_2)$ \cite{Landau,Ollivier98}. The phason dynamics is then governed by the following equations of motion:
\begin{subequations}\label{ec_mov_DPhG}\begin{align}
m \ddot \eta_2' +d^2\eta_2'-c\nabla^2\eta_2'+d\xi_2'&= 0,\\
\gamma_\xi \dot \xi_2' +\xi_2' + d\eta_2'&= 0.
\label{ec_mov_xi}
\end{align}\end{subequations}
Let us emphasize that the most important in these equations is not the extra variable $\xi_2$ by itself, but the frequency-dispersion that this variable yields in the phason dynamics. Strictly speaking, to reproduce the specific features of e.g. BCPS, i.e. gapped phasons and a central peak, it is only this dispersion what would be necessary (see below). However, to calculate the corresponding phason contribution to the specific heat, it is convenient to bear in mind that such a dispersion is obtainable from Eqs. \eqref{ec_mov_DPhG} (see Appendix). Anyway, the resulting phason dynamic is different at different time scales: while the slow oscillations are overdamped (central peak), the rapid ones are not (gapped phasons) \cite{Landau}. The phason contribution to the low-temperature specific heat then naturally split into two terms: $C = C_\text{gph}+C_\text{cp}$ (see Appendix for further details). The former:
\begin{align}
C_\text{gph}&=k_B\int {d^3k\over (2\pi)^3}
\left({\hbar \omega_2(k)\over 2k_BT}\right)^2
\left[\coth^2\left({\hbar \omega_2(k)\over 2k_BT}\right)-1\right]\nonumber \\
&= k_B {k_m^3 \over 2\pi^2}
\left({T \over \Theta}\right)^3\int_{x_0(T)}^{\Theta/T}dx
{[x^2-x_0^2(T)]^{1/2}x^3e^x\over (e^x-1)^2},
\label{C_ph}\end{align}
is the contribution that can be ascribed to the (undamped) gapped phasons \cite{note}. Here $\omega_2^2(k)=\delta^2+v^2k^2$, $\delta^2=d^2/m$,  $x_0(T)=\hbar\delta/(k_BT)$ and, assuming that $\delta \ll vk_m$, $\Theta\simeq \hbar vk_m/k_B$. The latter contribution:
\begin{align}
C_\text{cp}=k_B\int {d^3k\over (2\pi)^3} 
\bigg[&
\Big({\hbar \omega_\text{cp}(k)\over 2\pi k_BT}\Big)^2
\psi'\Big({\hbar \omega_\text{cp}(k)\over 2\pi k_BT}\Big)
\nonumber \\
&-
\Big({\hbar \omega_\text{cp}(k)\over 2\pi k_BT}\Big)-{1\over 2}
\bigg],
\label{C_cp}\end{align}
is due to damped phasons (central peak). Here $\omega_\text{cp}(k)=v^2k^2/(\delta^2\gamma_\xi/m)$ and $\psi'(x)={d\over dx}\big[{d\over dx}\ln \Gamma(x)\big]$ is the trigamma function. The low-$T$ asymptotic of this latter contribution is given by Eq. \eqref{low_T_specificheat_relax} by replacing $\gamma_\eta \to \gamma_\xi \delta^2$. 

Prior to estimate the phason contribution to the low-temperature specific heat in particular cases, let us see why this scheme of a dynamic phason gap is plausible in BCPS. (In short, this was already mentioned in Ref. \cite{Ollivier98}.) From Eqs. \eqref{ec_mov_DPhG} one can find the susceptibility associated with the phason:\begin{align}\chi_\text{ph}(k,\omega)=m^{-1}\big[\widetilde\omega_{2}^2(k,\omega)-\omega^2-i\gamma(\omega)\omega\big]^{-1},\label{chi}\end{align}where $\widetilde\omega_{2}^2(k,\omega)=\omega_{2}^2(k) -{\delta^2\over 1+\gamma_\xi^2 \omega^2}$, and $\gamma(\omega)={\gamma_\xi \delta^2\over 1+\gamma_\xi^2 \omega^2}$. According to this susceptibility, the spectrum of the inelastically scattered neutrons in the IC phase will show i) a central peak contribution and ii) a ``gapped phason'' contribution peaked at $\omega \simeq \delta$ if $\delta \gg \gamma_\xi^{-1}$. Although the central peak is not resolved by neutron scattering \cite{Ollivier98,Ollivier03}, the analysis of the corresponding intensities show that the situation in BCPS is compatible with this scenario \cite{Ollivier98}. In what concerns to NMR experiments, one has to bear in mind that NMR is a low-frequency probe. The corresponding Larmor frequency is then expected to be such that $\omega_L \ll \gamma_\xi^{-1}$ (it is actually the case in BCPS, see below). In consequence, the phason susceptibility probed in these experiments reduces to \begin{align}\chi_\text{ph}(k,\omega_L)\approx m^{-1}\big(v^2k^2-\omega_L^2-i\gamma_\xi\delta^2\omega_L\big)^{-1},\label{chi_NMR}\end{align}and, therefore, no gap will be revealed by NMR. (It can be said that only the central peak part of the spectrum of the inelastically scattered neutrons is probed by NMR.) It is worth mentioning that formula \eqref{chi_NMR} was used in fact in Ref. \cite{Souza91}. This further gives a $\omega_L^{1/2}$ dependence of the spin-lattice relaxation time, first derived in Ref. \cite{Zumer81}.

We are now in a position to estimate the phason contribution to the low-temperature specific heat in the IC phases of biphenyl and BCPS, and in the charge-density-wave system blue bronze $\rm K_{0.30}MoO_3$:

i) \emph{Biphenyl}. In accordance with NMR \cite{Liu85} and inelastic neutron scattering data \cite{Launois89}, the scheme of no phason gap seems to be valid here (the corresponding gap should be $\lesssim 10\, \rm MHz$ at least). From neutron scattering, the phason damping can be inferred such that $\gamma \sim 50 \,\rm GHz$. Notice that such a damping, obtained at $T=3\,\rm K$, is twelve orders of magnitude stronger than the one expected for a perfect crystal \cite{Levanyuk97a}. So defects must be the cause of the real damping, although inducing no (observable) phason gap. Both phason and sound velocities can be estimated as $\sim 10^{5}\, \rm cm\; s^{-1}$ \cite{Etrillard97,Launois89,Cailleau86}. Therefore, the linear-in-$T$ contribution due to phasons will prevail over the Debye one at temperatures $\lesssim 1\, \rm K$. In addition, according to Eq. \eqref{low_T_specificheat_relax}, it is obtained $C/T \sim 10\div 10^2~\rm erg \, cm^{-3} \, K^{-2}$ for these low temperatures. The experimental value $\sim 0.5 \, \rm erg \, cm^{-3} \, K^{-2}$ \cite{Etrillard97} is smaller in this case, what shall be commented below. 

ii) \emph{BCPS (dynamic phason gap)}. From NMR and inelastic neutron scattering experiments \cite{Souza91,Ollivier98,Ollivier03}, the corresponding damping can be inferred such that $\gamma_\xi \delta^2 \sim 10 \,\rm GHz$. Notice that no temperature dependence is observed in this damping down to $T=19\,\rm K$, so it seems very probable that it remains unaltered down to very low temperatures (as in biphenyl). According to Refs. \cite{Etrillard96,Cailleau86}, both phason and sound velocities are also $\sim 10^{5}\, \rm cm\, s^{-1}$. Then, the linear-in-$T$ contribution due to damped phasons is expected to be the leading one at temperatures $\lesssim 1 \rm \, K$. It would be such that $C/T \sim 10\div 10^2~\rm erg \, cm^{-3} \, K^{-2}$, what agrees in order of magnitude with the observed experimentally \cite{Etrillard96}. In this case (dynamic phason gap), in addition to this linear-in-$T$ contribution there is a contribution due to undamped gapped phasons [Eq. \eqref{C_ph}]. The phason gap can be estimated as $\delta \sim 100\, \rm GHz$ \cite{Ollivier98}, so the maximum in $C/T^3$ observed at $\sim 1\div 2\, \rm K$ can then be explained as a result of this latter contribution \cite{note}. This has already noticed in Ref. \cite{Etrillard96}.

iii) \emph{Blue bronze $ K_{0.30}MoO_3$}. Here, as well as in other charge-density-wave systems (see, e.g., Ref. \cite{Biljakovic91} and the references therein), the scheme of a static phason gap seems to be the most appropriate one. In accordance with inelastic scattering data \cite{Pouget91} this gap is $\sim 200\,\rm GHz$, and the phason damping can be estimated as $\gamma \sim 800\,\rm GHz$. Both phason \cite{Hennion92} and sound \cite{Dahlhauser86,Odin01} velocities can be inferred as $\sim 3\times 10^{5}\, \rm cm\; s^{-1}$. Consequently, a linear-in-$T$ contribution with $C/T \sim 10\div 10^2~\rm erg \, cm^{-3} \, K^{-2}$ is expected due to the damping of the phasons (prevailing over the Debye one at temperatures $\lesssim 1\, \rm K$) and, in addition, a maximum in $C/T^3$ at $T\simeq 10\rm \, K$ due to the phason gap. This qualitatively describes the experimental observations \cite{Dahlhauser86,Odin01}. Mention that in Ref. \cite{Lorenzo02} it is questioned the explanation that the maximum in $C/T^3$ can be ascribed to gapped phasons: it is shown that a similar maximum arises due to the acoustic anisotropy of this system. In Ref. \cite{Lasjaunias02} it is pointed out that, however, acoustic anisotropy cannot explain why this maximum is so sensitive to the crystal quality. It is also worth mentioning that, given the relatively large anisotropy in the phason velocity \cite{Hennion92}, phasons can behave as overdamped oscillations along certain directions. Bearing in mind that an overdamped phason branch in the one-dimensional case give rise to a contribution $\sim T^{1/2}$ in the low-temperature specific heat \cite{Iengo98}, this could explain the deviation of the linear-in-$T$ behavior observed experimentally in Ref. \cite{Odin01} at the lowest temperatures. 

It is worth mentioning that, at present, the above contributions to the low-temperature specific heat due to the damping of the phason can only be estimated in order of magnitude. It is true that the corresponding phason dynamics has been extensively studied in a number of papers. However, the quantitative comparison of our results with the experimental data would require, in particular, the knowledge of the phason damping for the whole phason branch. Unfortunately, these latter data are not reported in the literature and consequently, when carrying out the comparison of our results with experimental data, we are forced to make some assumptions. For instance that the phason damping is the same for the whole phason branch. These type of assumptions, at the present unavoidable, could be the reason for the overestimation of the linear-in-$T$ contribution due the phason damping when comparing with experimental observations. More precise estimates require more complete experimental data (see Ref. \cite{Cano/cond-mat/0404063} for a detailed discussion). Nevertheless, the agreement in order of magnitude we have obtained indicates that the deviations from the Debye law observed in the very-low-temperature specific heat of IC phases are, the most probably, in regard to the phason damping. Mention also that the resulting linear-in-$T$ behavior is expected as a low-temperature asymptotic limit. It is well possible that in the above mentioned experiments this limit has not been fully achieved but rather are in an intermetiate (crossover) region, which could be the reason of the different power laws observed there. 

In summary, we have shown that the glass-like anomaly observed in the low-temperature specific heat of incommensurate phases can be explained as a result of the phason damping. Three possible scenarios for the phason dynamic, reproducing the corresponding NMR and inelastic scattering data for biphenyl, BCPS and blue bronze $\rm K_{0.30}MoO_3$, are discussed. Estimates of the corresponding low-temperature specific heat are in reasonable agreement with the observed experimentally.

We thank K. Biljakovi$\rm \acute c$ for useful discussions.

\appendix

\section{Phason contribution to the specific heat}

The knowledge of the partition function $Z$ of a system permits the computation of the thermodynamic magnitudes. We are interested, in particular, in the specific heat: $C = \beta^{-1}[\partial^2 (T\ln Z) /\partial T^2]$, where $\beta=1/(k_BT)$.

Formally, $Z$ can be written as the path integral \cite{Feynman_Hibbs,Negele_Orland,Simanek,Peskin_Schroeder,Weiss}
\begin{align}
Z= \oint {\cal D}[\phi_i({\mathbf r},\tau)]\exp\left(-S[\phi_i]/\hbar\right), 
\label{Z_total}\end{align}
where $\phi_i({\mathbf r},\tau)$ (${\mathbf r}\in V$, $\tau \in [0,\hbar \beta]$) are the fields characterizing the system, $S[\phi_i]$ is the Euclidean (imaginary-time) action of the system, and the path sum is over the fields satisfying the appropriate boundary conditions. In what concerns to the $\hbar \beta$-space of the (imaginary) time variable $\tau$, these boundary conditions are periodic ones for fields associated with bosons (to our purposes, it is sufficient to consider here bosonic systems). 

Among the fields characterizing the system, it can be identified the one associated with the degrees of freedom of interest, say $\eta({\mathbf r},\tau)$. Formal integration over all other degrees of freedom, i.e., over the set of fields $\{\varphi_i({\mathbf r},\tau)\}$, gives 
\begin{align}
Z=\oint {\cal D}[\eta({\mathbf r},\tau)]
\exp\left(-\widetilde S_\text{eff}[\eta]/\hbar\right), 
\label{}\end{align}
where the effective Euclidean action $\widetilde S_\text{eff}[\eta]$ is such that
\begin{align}
\exp\left(-\widetilde S_\text{eff}[\eta]/\hbar\right)
=\oint {\cal D}[\varphi_i({\mathbf r},\tau)]
\exp\left(-S[\eta,\varphi_i]/\hbar\right). 
\label{}\end{align}
This action can be divided into $\widetilde S_\text{eff}[\eta]=S_\text{eff}^{(0)} + S_\text{eff}[\eta]$. Thus the partition function can be expressed as $Z=Z_0 Z_\text{red}$, where $Z_0 =\exp\big(-S_\text{eff}^{(0)}/\hbar\big)$ and
\begin{align}
Z_\text{red}=\oint {\cal D}[\eta({\mathbf r},\tau)]\exp\left(-S_\text{eff}[\eta]/\hbar\right). 
\label{Z_red}\end{align}
The reduced partition function $Z_\text{red}$ results from the contribution due to the subsystem characterized by $\eta({\mathbf r},\tau)$. In contrast to the action in Eq. \eqref{Z_total}, the effective action $S_\text{eff}[\eta]$ in Eq. \eqref{Z_red} may be, e.g., a temperature-dependent functional. It is because $S_\text{eff}[\eta]$ arises from a partial integration. This procedure is well known, e.g., in the theory of phase transitions (see, e.g., Ref. \cite{Strukov_Levanyuk} for an exposure in the classical case). It is worth mentioning that $S_\text{eff}[\eta]$ has not to be hermitian necessarily. Relaxational processes can then be described within this formalism. We are interested, in particular, in the contribution to the specific heat due to phasons (damped oscillators). Let us illustrate the method that further we shall use to calculate such a contribution considering first the case of ``normal'' phonons (undamped oscillators). 

\subsection{Noninteracting undamped phonons: Path-integral calculation of the partition function} 

In the simplest case, the effective Euclidean action for phonons can be taken as
\begin{align}
S_\text{eff}[\eta] =\int d{\mathbf r}\int_0^{\hbar \beta} d\tau \left(
{m\over 2} (\partial_\tau\eta)^2 + {a\over 2}\eta^2 + {c\over 2}(\nabla \eta)^2
\right),
\label{S_eff}\end{align}
where $\eta({\mathbf r},\tau)$ is a (bosonic) displacement field. The first term of this action is associated with the kinetic energy while the last two terms do with the potential one. The case $a=0$ corresponds to acoustic phonons while $a\not =0$ does with optic ones. 

The first variation of the action gives the corresponding equation of motion. According to Eq. \eqref{S_eff} it reads
\begin{align}
m\Lambda \eta= 
m\left(-\partial_\tau^2 + \omega_0^2 - v^2 \nabla^2\right)\eta=0
\label{equ_mot_hb}\end{align}
in the $\hbar\beta$-space, where $\omega_0^2=a/m$, $v^2 =c/m$, and $\Lambda$ is the so-called fluctuation operator (the reason of such a designation can be seen below). 

By putting $\eta= \eta_0+\eta'$ in Eq. \eqref{S_eff}, with $\eta_0$ satisfying the corresponding equation of motion together with periodic boundary conditions, and integrating by parts, we find that
\begin{align}
S_\text{eff}[\eta] =S_\text{eff}[\eta_0]+ {m\over 2}
\int d{\mathbf r}\int_0^{\hbar \beta} d\tau \eta'\Lambda
\eta'
\label{}\end{align}
for periodic fluctuations $\eta'$. Consequently, the reduced partition function \eqref{Z_red} can be expressed as $Z_\text{red}= Z_\text{red}^{\circ} Z_\text{red}^\text{fl}$, where $Z_\text{red}^{\circ}=\exp\left(-S_\text{eff}[\eta_0]/\hbar\right)$ and $Z_\text{red}^\text{fl}$ is associated with fluctuations:
\begin{align}
Z_\text{red}^\text{fl}&=\oint {\cal D}[\eta'({\mathbf r},\tau)]\exp\left(-{m\over 2\hbar}\int d{\mathbf r}\int_0^{\hbar \beta}d\tau 
\eta'\Lambda
\eta'\right)
\label{}\end{align}
(Gaussian fluctuations in our case). 

The only solution of Eq. \eqref{equ_mot_hb} which is periodic with respect to the $\tau$ variable reduces to $\eta_0=0$. So that $Z_\text{red}^{\circ}=1$ in our case. Expressing $\eta'$ as the Fourier transform
\begin{align}
\eta'({\mathbf r},\tau)=\sum_{{\mathbf k}=0}
\sum_{n=-\infty}^{\infty} 
\eta'_{{\mathbf k},\nu_n} e^{i{\mathbf k}\cdot {\mathbf r}}e^{i{\nu_n}\tau},
\end{align}
where $\nu_n=2\pi n/(\hbar \beta)$ are the so-called bosonic Matsubara frequencies \cite{nota_A1}; and writing down the functional measure as 
\cite{note_measure}
\begin{widetext}
\begin{align}
\oint {\cal D}[\eta'({\mathbf r},\tau)]\dots \equiv 
\prod_{{\mathbf k}=0}
\left[
\int{d\eta'_{{\mathbf k},0}\over \sqrt{2\pi\hbar^2\beta/m}}
\prod_{n=1}^{\infty}\left(\iint 
{d\text{Re }\eta'_{{\mathbf k},\nu_n}d\text{Im }\eta'_{{\mathbf k},\nu_n} 
\over \pi/(m\beta\nu_n^2)}\right)\right]\dots,
\label{measure_factor}\end{align}
we have
\begin{align}
Z_\text{red}
&=\prod_{{\mathbf k}=0}
\left[\int{d\eta'_{{\mathbf k},0}\over \sqrt{2\pi\hbar^2\beta/m}}
\prod_{n=1}^{\infty}\iint\left( 
{d\text{Re }\eta'_{{\mathbf k},\nu_n}d\text{Im }\eta'_{{\mathbf k},\nu_n} 
\over \pi/(m\beta\nu_n^2)}\right)\right]
\exp\left(-{m\beta\over 2}\sum_{{\mathbf k}=0}
\sum_{n=0}^{\infty}
\hat\Lambda (k,\nu_n) 
|\eta'_{{\mathbf k},\nu_n}|^2\right)
\nonumber \\
&=\prod_{{\mathbf k}=0}
\left({1\over \hbar\beta \sqrt{\hat \Lambda (k,0) }} 
\prod_{n=1}^{\infty}
{\nu_n^2 \over \hat \Lambda(k,\nu_n)}\right)
=\prod_{{\mathbf k}=0}
\left({1\over \hbar\beta \omega_0(k)} 
\prod_{n=1}^{\infty}
{\nu_n^2 \over \omega_0^2(k)+\nu_n^2}\right)
\nonumber \\
&=\prod_{{\mathbf k}=0}
{1\over 2\sinh[\hbar\beta\omega_0(k)/2]},
\label{}\end{align}
where $
\hat\Lambda(k,\nu_n)=\omega_0^2(k) + \nu_n^2$ and $\omega_0^2(k)=\omega_0^2 + v^2k^2$ (note that the reduced partition function we obtain is exact because it is connected to a quadratic effective action). In the last form, we have the usual expression for the partition function of a system of noninteracting undamped phonons (see, e.g., Ref. \cite{Feynman_Hibbs}).

From Eq. \eqref{equ_mot_hb} it can be seen that the eigenvalues of the operator $\Lambda$ gives the inverse of the (linear) dynamic generalized susceptibility: $\chi^{-1}({\mathbf k} ,\omega)=m\Lambda ({\mathbf k},\omega)$. 
When acting on $\hbar\beta$-periodic functions, these eigenvalues can be obtained by the transformation
\begin{align}
\Lambda ({\mathbf k} ,\omega) \quad\longrightarrow\quad
\Lambda ({\mathbf k} ,i\nu_n)\equiv \hat \Lambda ({\mathbf k},\nu_n).
\label{}\end{align}
[In the case of phonons: $\Lambda ({\mathbf k},\omega)=\omega_0^2(k)- \omega^2 
\longrightarrow
\Lambda ({\mathbf k} ,i\nu_n)\equiv \hat \Lambda ({\mathbf k},\nu_n)
=\omega_0^2(k) + \nu_n^2$.]
As we have seen, the knowledge of these latter eigenvalues permits the computation of the reduced partition function. It is the case even for relaxational dynamics. In the following, we shall take advantage of this point when calculating the contribution to the specific heat associated with phasons.

\subsection{Phason contribution to the specific heat} 

\subsubsection{Scheme of no phason gap}

Let us now calculate the phason contribution to the specific heat within the scheme of no phason gap. The procedure we shall follow is based on the results concerning to the damped harmonic oscillator, which can be found, e.g., in Refs. \cite{Simanek,Weiss}. Formulas we generate are analogous to that given in Ref. \cite{Iengo98}.

According to Eq. \eqref{eq_mot_eta_2} the phason susceptibility is 
\begin{align}
\chi_\text{ph}(k,\omega)
={1\over m\big[\omega_{2}^2(k)-\omega^2-i\gamma\omega\big]},
\label{}\end{align}
where $\omega_{2}^2(k)= v^2k^2$, $v^2=c/m$, and $\gamma=\gamma_\eta/m$. From this susceptibility one finds that
\begin{align}
\Lambda_\text{ph}(k,\omega)&\equiv [m\chi_\text{ph}(k,\omega)]^{-1}\nonumber\\
&=
{[\omega+i\lambda_1(k)][\omega+i\lambda_2(k)][\omega+i\lambda_3(k)]
\over \omega+i\omega_D},
\label{Lambda}\end{align}
where 
\begin{subequations}\label{relations}\begin{align}
\lambda_1(k)+\lambda_2(k)+\lambda_3(k)&= \omega_D, \\
\lambda_1(k)\lambda_2(k)+ \lambda_2(k)\lambda_3(k)+\lambda_1(k)\lambda_3(k)
&=\omega_2^2(k)+\gamma \omega_D,\\
\lambda_1(k)\lambda_2(k)\lambda_3(k)&=\omega_D\omega_2^2(k),
\end{align}\end{subequations} 
and $\omega_D^{-1}$ is a memory time associated with a Drude regularization of the damping term in Eq. \eqref{eq_mot_eta_2}:
\begin{align}
\gamma_\eta \dot \eta_2(t) \to \gamma_\eta  \omega_D \int_{-\infty}^t dt'e^{-\omega_D(t-t')}\dot \eta_2(t').
\end{align}
Below we shall take the limit $\omega_D\to \infty$ in which this regularized damping term coincides with the original one. But, at this start point, such a regularization is necessary in order that the infinite-product expression for the partition function of the system, see Eq. \eqref{parti_nogap} below, actually exists (see Ref. \cite{Weiss,Weiss84,Iengo98}).

The phason contribution to the partition function of the system can be computed from $\Lambda_\text{ph}$ by going to the space of $\hbar \beta$-periodic functions, i.e., by replacing $\omega \to i\nu_n$ with $\nu_n =2\pi n/(\hbar \beta)$ \cite{Simanek,Weiss} (as we have seen, $\Lambda$ then gives the eigenvalues of the so-called fluctuation operator \cite{Weiss}):
\begin{align}
Z_\text{ph}&=\prod_{{\mathbf k}=0}\left(
{1\over \hbar \beta \sqrt{\Lambda_\text{ph}(k,0)}}
\prod_{n=1}^{\infty}{\nu_n^2 \over \Lambda_\text{ph}(k,i\nu_n)}\right)
=
\prod_{{\mathbf k}=0}\left(
{1\over \hbar \beta \omega_2(k)}
\prod_{n=1}^{\infty}{\nu_n^2 (\nu_n+\omega_D)
\over [\nu_n+\lambda_1(k)][\nu_n+\lambda_2(k)][\nu_n+\lambda_3(k)]}\right)
\nonumber \\&
=
\prod_{{\mathbf k}=0}\left(
{\hbar\beta \Big({\lambda_1(k)\lambda_2(k)\lambda_3(k)\over\omega_D}\Big)^{1/2}\over 4\pi^2}
{\Gamma \big[\lambda_1(k)/\nu\big] \Gamma \big[\lambda_2(k)/\nu\big] \Gamma \big[\lambda_3(k)/\nu\big] \over \Gamma \big(\omega_D/\nu\big)}\right),
\label{parti_nogap}\end{align}
where $\nu =\nu_1$ and $\Gamma(x)$ is the gamma function. We then have that the corresponding contribution to the specific heat is 
\begin{align}
C =
k_BT{\partial ^2 (T \ln Z_\text{ph}) \over \partial T^2}
=k_B \int {d^3k\over (2\pi)^3}\bigg\{
\sum_{i=1}^3[\lambda_i(k)/\nu]^2 \psi'[\lambda_i(k)/\nu]
-(\omega_D/\nu)^2 \psi'(\omega_D/\nu)-1\bigg\},
\label{C_exact}\end{align}
where $\psi'(x)=d^2[\ln \Gamma(x)]/dx^2$ is the trigamma function and summation over wavevectors has been replaced by integration.

By inserting the asymptotic expression $x^2 \psi'(x)\underset{x\to \infty}{\approx} x + 1/2+1/(6x)$ into the integrand in Eq. \eqref{C_exact} and using the relations \eqref{relations}, we further get the asymptotic low-temperature specific heat due to phasons:\begin{align}
C \underset{T\to 0}{\approx} {k_m^3\over 6\pi}{k_B^2T\over \hbar (v^2k_m^2/\gamma)}.
\label{cv_cp}\end{align}

\subsubsection{Scheme of dynamic phason gap}

Let us finally calculate the phason contribution to the specific heat within the scheme of a dynamic phason gap. Instead of Eq. \eqref{ec_mov_xi}, let us consider the equation
\begin{align}
\mu \ddot \xi_2(t) +\int_{-\infty}^{t}dt'\gamma_\xi(t-t')\dot \xi_2(t')+\xi_2(t)+d\eta_2(t)=0, 
\label{ec_mov_xi_regu}\end{align}
where $\gamma_\xi (t)=\gamma_\xi \omega_D\exp (-\omega_D t)$. Here we have introduced a inertial term and considered a Drude regularization for the damping one. Similar to the case of no phason gap, below we take the limit $\mu \to 0$ and $\omega_D\to \infty$ in which Eq. \eqref{ec_mov_xi_regu} reduces to Eq. \eqref{ec_mov_xi}. 

From the corresponding susceptibilities we get 
\begin{align}
\widetilde \Lambda_\text{ph}(k,\omega) &\equiv 
m\mu\big[\chi_{\eta_2\eta_2}^{-1}(k,\omega)\chi_{\xi_2\xi_2}^{-1}(k,\omega) -\chi_{\eta_2\xi_2}^{-2}(k,\omega) \big]
=[\omega_2^2(k)-\omega^2]
\Big(w^2-\omega^2-iw^2{\gamma\omega_D\omega\over \omega_D - i\omega}\Big)-\delta^2w^2
\nonumber \\ &
=
{[\omega+i\lambda_1(k)][\omega+i\lambda_2(k)][\omega+i\lambda_3(k)]
[\omega+i\lambda_4(k)][\omega+i\lambda_5(k)]
\over 
\omega + i\omega_D},
\end{align}
where $\omega_2^2=\delta^2+v^2k^2$, $w^2=1/\mu$ and now $\gamma = \gamma_\xi /m$. 

Within the scheme of the dynamic phason gap we are considering, the phason contribution to the partition function of the system must be understood as the contribution of the degrees of freedom represented by $\eta_2$ and $\xi_2$, i.e., the contribution of linearly coupled oscillators. Following Ref. \cite{Simanek,Weiss} one can see that it can be computed from $\widetilde \Lambda_\text{ph}$:
\begin{align}
Z_\text{ph}&=
\prod_{{\mathbf k}=0}\left(
{1\over (\hbar \beta)^2 \sqrt{\widetilde \Lambda_\text{ph}(k,0)}}
\prod_{n=1}^{\infty}{\nu_n^4 \over \widetilde\Lambda_\text{ph}(k,i\nu_n)}
\right)
\nonumber \\
&=
\prod_{{\mathbf k}=0}\left(
{1\over (\hbar \beta)^2 w\sqrt{\omega_2^2(k)-\delta^2}}
\prod_{n=1}^{\infty}{\nu_n^4 (\nu_n+\omega_D)
\over [\nu_n+\lambda_1(k)][\nu_n+\lambda_2(k)][\nu_n+\lambda_3(k)]
[\nu_n+\lambda_4(k)][\nu_n+\lambda_5(k)]}
\right)
\nonumber \\
&=
\prod_{{\mathbf k}=0}\left(
{(\hbar \beta)^2 \left(\lambda_1(k)\lambda_2(k)\lambda_3(k)\lambda_4(k)\lambda_5(k)\over \omega _{D}\right)^{1/2}\over 2^4\pi^4}
{\Gamma[\lambda_1(k)/\nu]\Gamma[\lambda_2(k)/\nu]\Gamma[\lambda_3(k)/\nu] \Gamma[\lambda_4(k)/\nu]\Gamma[\lambda_5(k)/\nu]
\over \Gamma(\omega_D/\nu)}
\right).
\label{}\end{align}
We then have that the phason contribution to the specific heat is 
\begin{align}
C &=k_BT{\partial ^2 (T \ln Z_\text{ph}) \over \partial T^2}
=
k_B\int{d^3k\over (2\pi)^3}\bigg\{\sum_{i=1}^5[\lambda_i(k)/\nu]^2 \psi'[\lambda_i(k)/\nu]
-(\omega_D/\nu)^2 \psi'(\omega_D/\nu)-2\bigg\}.
\label{specific_heat_dynamic}\end{align}

We are interested in the limiting case $\omega_2^2(k)-\delta^2,\gamma^2 ,\omega_2^2(k) \ll w^2 \ll\omega_D^2$. We then have
\begin{align}
\lambda_1(k)\approx \omega_\text{cp}(k),\qquad
\lambda_2(k)\approx \gamma w^2-\omega_\text{cp}(k),\qquad 
\lambda_3\approx \omega_D - \gamma w^2,\qquad
\lambda_{4,5}(k)\approx\pm i\omega_2(k),
\end{align}
where $\omega_\text{cp}(k) ={v^2k^2/(\delta^2\gamma)}$. A a result, in Eq. \eqref{specific_heat_dynamic} the contribution due to the central peak (damped phasons) and the one due to the (undamped) gapped phasons naturally split up: $C = C_\text{cp}+C_\text{gph}$. The low-$T$ asymptotic of the former contribution coincides with Eq. \eqref{cv_cp} by replacing $\gamma_\eta \to \gamma_\xi\delta^2$. The latter is given by
\begin{align}
C_\text{gph}&=
k_B\int{d^3k\over (2\pi)^3}
\Big\{
[\lambda_1(k)/\nu]^2\psi'[\lambda_1(k)/\nu]
+[\lambda_2(k)/\nu]^2\psi'[\lambda_2(k)/\nu]-1\Big\}
\nonumber \\ &
\approx-
k_B\int{d^3k\over (2\pi)^3}
\Big\{
1+2[\omega_2(k)/\nu]^2 
\text{Re }\psi'[i\omega_2(k)/\nu]\Big\}
=
k_B\int{d^3k\over (2\pi)^3}
\Big({\pi\omega_2(k)\over \nu}\Big)^2\Big[\coth^2\Big({\pi\omega_2(k)\over\nu}\Big)-1\Big]
\nonumber \\
&=k_B {k_m^3 \over 2\pi^2}
\left({T \over \Theta}\right)^3\int_{x_0(T)}^{\Theta/T}dx
{[x^2-x_0^2(T)]^{1/2}x^3e^x\over (e^x-1)^2},
\end{align}
where $x_0(T)=\hbar\delta/(k_BT)$ and $\Theta\simeq \hbar vk_m/k_B$ (here it has been assumed that $\delta \ll vk_m$).

\vspace{5pt}
\end{widetext}


\begin{references}
\bibitem{nota_A1} We are considering real periodic functions with respect to the space coordinates and the imaginary-time variable $\tau$, so that $\eta'_{{\mathbf k},\nu_n}=\eta'^*_{-{\mathbf k},-\nu_n}$.
\bibitem{note_measure} One could expect that such a measure merely be 
\begin{align}
\oint {\cal D}[\eta'({\mathbf r},\tau)]\dots \equiv 
\prod_{{\mathbf k}=0}
\left[
\int{d\eta'_{{\mathbf k},0}}\prod_{n=1}^{\infty}\left(\iint 
{d\,\text{Re}\,\eta'_{{\mathbf k},\nu_n}d\,\text{Im}\,\eta'_{{\mathbf k},\nu_n} 
}\right)\right]\dots
\nonumber \label{}\end{align}
The origin of the nontrivial factor appearing in Eq. \eqref{measure_factor} can be understood following, e.g., Ref. \cite{Peskin_Schroeder}. The starting point should be, in fact, to consider the effective Euclidean action 
\begin{align}
\widetilde{S}_\text{eff}[\eta,\pi]=\int d{\mathbf r}\int_0^{\hbar \beta} 
d\tau \left(
{a\over 2}\eta^2 + {c\over 2}(\nabla \eta)^2
+{1\over 2m} \pi^2 -i\pi (\partial_\tau \eta)
\right),\nonumber 
\label{}\end{align}
instead of that given in Eq. \eqref{S_eff}. Thus, in addttion to the field $\eta({\mathbf r},\tau)$ associated with the displacement amplitudes, we take into account the one representing its conjugated momentum density $\pi({\mathbf r},\tau)$. Since this action is quadratic in $\pi({\mathbf r},\tau)$, we can complete the square and further evaluate the path integral over $\pi({\mathbf r},\tau)$ in the reduced partition function:
\begin{align}
Z_\text{red}&=\oint 
\widetilde{\cal D}[\eta({\mathbf r},\tau)]\widetilde{\cal D}[\pi({\mathbf r},\tau)]
\exp\left(-\widetilde S_\text{eff}[\eta,\pi]/\hbar\right)=\oint 
{\cal D}[\eta({\mathbf r},\tau)]
\exp\left(-S_\text{eff}[\eta]/\hbar\right),\nonumber 
\label{}\end{align}
where $S_\text{eff}[\eta]$ coincides with \eqref{S_eff}
and the new measure ${\cal D}[\eta({\mathbf r},\tau)]$ is the old one $\widetilde{\cal D}[\eta({\mathbf r},\tau)]$ multiplied by mentioned factor.
\end{references}
\end{document}